\documentclass[twocolumn,prl,aps,epsf,showpacs]{revtex4}

\usepackage{amssymb,epsf}
\usepackage{graphicx}
\usepackage{epstopdf}

\begin{document}

\title{Evidence for Skyrmion Crystallization from NMR Relaxation Experiments}

\author{G. Gervais$^{1,2}$, H.L. Stormer$^{1,3}$, D.C. Tsui$^{4}$, P.L. Kuhns$^{2}$, W.G. Moulton$^{2}$, A.P. Reyes$^{2}$, L.N. Pfeiffer$^{3}$, K. W. Baldwin$^{3}$, and K.W. West$^{3}$}

\address{$^{1}$Department of Physics and Department of Applied Physics, Columbia University, New York, NY 10027 USA\\
$^{2}$National High Magnetic Field Laboratory, Tallahassee, FL 32306, USA\\
$^{3}$Bell Laboratories, Lucent Technology, Murray Hill, NJ 07974 USA\\
$^{4}$Department of Electrical Engineering, Princeton University
Princeton, NJ 08544 USA\\}

\begin{abstract} 
A resistively detected NMR technique was used to probe the two-dimensional
electron gas in a GaAs/AlGaAs quantum well. The spin-lattice relaxation rate
$(1/T_{1})$ was extracted at near complete filling of the first Landau level by
electrons. The nuclear spin of $^{75}$As is found to relax much more efficiently
with $T\rightarrow 0$ and when a well developed quantum Hall state with
$R_{xx}\simeq 0$ occurs. The data show a remarkable correlation between
the nuclear spin relaxation and localization. This suggests that the magnetic
ground state near complete filling of the first Landau level may contain a
lattice of topological spin texture, i.e. a Skyrmion crystal. 

\end{abstract}

\maketitle
\narrowtext

In his seminal work on nuclear matter more than forty
years ago Skyrme  showed that baryons emerge mathematically as a 
static solution of a meson field described by the so-called Skyrme Lagrangian \cite{Skyrme62}. 
His work provided the foundation for the quantum theory of solitons, and 
more recently found an interesting and {\it a priori} surprising connection to the
physics of electrons confined to a two-dimensional plane.
In the presence of a strong perpendicular magnetic field, the
orbital motion of these electrons is quantized into discrete Landau levels.
When only the lowest of such level
is almost completely occupied, the elementary excitations of the system become large topologically
stable spin texture known as Skyrmions \cite{Sondhi93}. It was further proposed
that at T=0 Skyrmions would localize on a square lattice \cite{Brey95}. This
ground state represents a new type of magnetic
ordering which possesses long-range orientation and positional order.
This is the solid-state analogue of the Skyrmion crystal state which 
is used to describe dense nuclear matter using Skyrme's topological 
excitation model\cite{Klebanov85}.
In this work, we present an extensive study of nuclear magnetic resonance (NMR) spin-lattice relaxation rate in the first Landau level
of  an extremely high-quality GaAs/AlGaAs sample. We find strong and enhanced relaxation 
in the limit of $T\rightarrow 0$ and $R_{xx}\rightarrow 0$ where localization of electronic states
occurs. This is consistent with previous measurements of the heat capacity
in multiple quantum wells at very low temperatures \cite{Bayot96,Bayot97}, and 
with the predictions of a magnetic ground state containing a Skyrme crystal\cite{Cote97}.

Several theoretical publications \cite{Sondhi93,Brey95,Cote97,Fertig94,Fertig97,Timm98,Green00,Sinova00} 
have pointed toward 
the existence of Skyrmions in a two-dimensional electron gas (2DEG). 
At filling factor $\nu= 1$, where $\nu$ is defined by the ratio of
 the electronic density  $n$ to the magnetic flux density, $\nu=\frac{n}{B/\Phi}=\frac{nh}{eB}$, 
  the quantized Hall state is ferromagnetic. For sufficiently small Zeeman-to-Coulomb 
  energy ratio $\eta=E_{z}/E_{c}=\frac{g^{\star}\mu_{B} B}{e^{2}/\epsilon l_{B}}$, 
  where $g^{\star}$ is the electronic g-factor and $l_{B}=\sqrt{\hbar/eB}$ is the magnetic length, 
   Sondhi {\it et al.} showed that  the low-lying excitations
  are not single spin-flips, but rather a smooth distortion of the spin
  field in which several spins (4-30) participate \cite{Sondhi93}. These Skyrmions are topologically stable, charged $\pm e$, and gapped excitations which are 
  the result of an energy tradeoff where a higher Zeeman cost is paid for the profit of lowering the 
exchange energy between neighboring spins. While several theoretical 
proposals suggest a lattice state of Skyrmions (\cite{Brey95,Timm98,Green00})
it is theoretically debated\cite{Paredes99} whether
this is the case or whether a  `liquid' state prevails.

Previous measurements of the
electronic spin polarization by NMR around $\nu =1$ by Barrett {\it et al.} 
showed strong evidence for a finite density of Skyrmions
at temperature $T\sim 1 K$ \cite{Barrett95,Tycko95}. Subsequently, 
Schmeller {\it et al.}\cite{Schmeller95} used tilted-transport measurements to 
show that as many as seven electron spins 
participate in the spin excitations at $\nu=1$, while at all other integer fillings only
single spin-flip excitations were observed. Other experiments using various experimental probes 
confirmed such findings and expanded on them \cite{Melinte01,Khandelwal01, Groshaus04}. 
Measurements of the heat capacity
in multiple quantum wells showed a sharp heat capacity
peak at $T\sim 40$ mk near $\nu\sim 1$ which was interpreted as a possible transition between
a liquid and lattice state of skyrmions \cite{Bayot96,Bayot97}.
Our experiment adresses
the limiting $T\rightarrow 0$ behavior near $\nu \sim 1$  by measuring the nuclear spin-lattice relaxation 
rate $(1/T_{1})$ using  resistively detected NMR\cite{Dobers88,Kronmuller99,Smet02,Desrat02,Hashimoto02,Stern04}. 
Recently, Desrat {\it et al.} \cite{Desrat02}
exploited this technique, and
observed resistively detected NMR over a broad range of filling factors. 
 In particular, they found an `anomalous' dispersive-like line shape near $\nu\sim 1$,
 for which they speculated as originating from a possible coupling to a Skyrme crystal.

The experiment was performed on a 40 nm wide  modulation-doped GaAs/AlGaAs
quantum well grown by molecular beam epitaxy. 
The electron density of the 2DEG was determined 
to be $n=1.60(1)\times 10^{11}$ $cm^{-2}$, and the mobility $\mu\simeq 17\times 10^{6}$ $cm^{2}/V\cdot s$. Cooling of the electrons down to temperatures near the base temperature ($\sim20$ mK) was achieved
by thermally anchoring the sample leads on the 
refrigerator by means of various powder and RC filters with low cutoff frequencies. 
The temperature was determined with a ruthenium oxide thermometer. 
 Figure 1 shows an example of magnetotransport measurement 
  at $T\simeq 20$ mK.

 \begin{figure}
\begin{center}
\includegraphics[width=8.5cm]{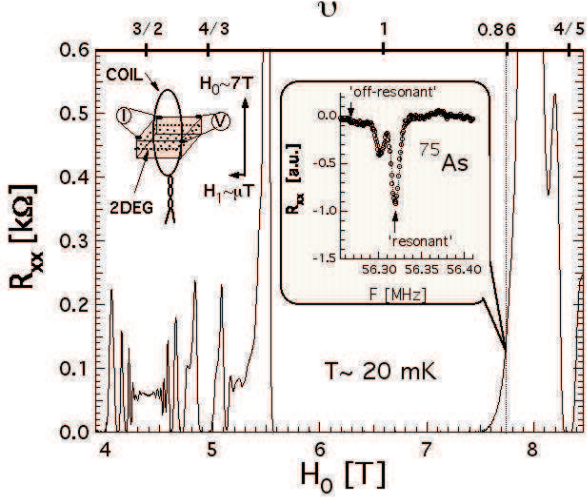}
\end{center}
\vspace*{-5mm}\caption{  {\small  Magnetotransport $R_{xx}$ for the two-dimensional electron
system (2DES). 
The bottom x-axis shows the applied magnetic field and the upper x-axis the corresponding
filling factors. A schematic of the NMR experiment is shown in the upper left corner. 
The inset at the centre of the graph shows an example of 
resistively detected NMR resonance for the $^{75}$As nucleus and at filling factor $\nu=0.86$}.}
\label{fig3}
\end{figure}

A cartoon of our experiment is shown in Fig.1. 
An NMR coil  is wrapped around the sample which resides in a strong ($\sim$ 8T)
perpendicular field $H_{0}$. A small radio-frequency (RF)  field $|H_{1}cos(\omega t)|\sim \mu T$
matching the NMR frequency $f_{NMR}=\gamma H_{0}$ is radiated on the sample,
 where $\gamma =7.29$ MHz/T for the $^{75}$As nuclei.
The resistance, $R_{xx}$, is monitored at constant $H_{0}$ and $T$
 while the RF-field if slowly swept across the nuclear resonance 
at a rate of $\sim$ 0.15 kHz/s.
A typical resistively detected NMR signal 
 for the $^{75}$As nucleus is shown in the central inset of Fig.1 at filling
 factor $\nu=0.86$.
  A small, but sizable resistance change $\delta R_{xx}$ is 
observed at resonance having typical signal strength $\delta R_{xx}/R_{xx}\sim 1\%$.
The detection scheme in resistive NMR relies on the hyperfine interaction ${\cal A}{\vec I}\cdot{\vec S}$.
 In GaAs, the electronic Zeeman energy for the 2DEG can be written as
 $E_{z}=g^{\star}\mu_{B}(H_{0}+B_{N})S_{z}$, where $B_{N}={\cal A}<I_{z}>/g^{\star}\mu_{B}$ 
 is the Overhauser shift,  ${\cal A}$ the hyperfine constant and  $<I_{z}>$ the nuclear
 spin polarization.
 Applying an RF-field at resonance modifies the thermal distribution of the 
nuclear spins in the applied field $H_{0}$ which depolarizes
the nuclear spins and as a consequence modifies the Zeeman gap, $\Delta_{z}=g^{\star}\mu_{B}(H_{0}+B_{N})$.
 For odd integer filling factors such as near $\nu\sim 1$ , the electronic transport in the thermally activated regime 
 is given by  $R_{xx}\sim e^{\frac{-\Delta_{z}}{2k_{B}T}}$  which makes possible 
 the detection of the NMR by means of resistivity. A study and a discussion of the lineshape will be published elsewhere\cite{Gervais04}.
 

 \begin{figure}
\begin{center}
\includegraphics[width=7.cm]{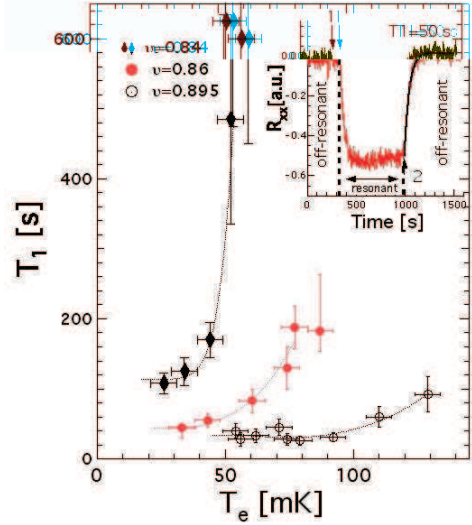}
\end{center}
\vspace*{-5mm}
\caption{  {\small  Nuclear spin-lattice relaxation time $T_{1}$
for $^{75}$As  at various electronic temperatures $T_{e}$
and filling factors, $\nu$. The thin dotted lines are guides-to-the-eyes. The inset shows a typical 
experiment from which $T_{1}$ is determined  (see text). The solid line is a fit to a single 
exponential.}}

\end{figure}

The inset of Fig.2 shows an example of 
spin-lattice relaxation time $T_{1}$ measurements. 
The resistance of  the 2DEG is initially 
measured at constant field $H_{0}$ and
temperature T and monitored in time
with the frequency set to be off-resonance (see the arrows in the central inset of Fig.1). 
At the time labeled `1', the frequency 
is moved on-resonance (see  Fig.1) and the resistance decreases 
as a consequence of  the nuclei being depolarized, and 
eventually reaches a steady state. At   `2', the frequency is 
set back to off-resonance, and the resistance decays
to its original state as the macroscopic nuclear magnetization ${\cal M}(t)$ relaxes
in a time $T_{1}$ to its thermal equilibrium value, ${\cal M}_{0}$. The time dependence of $R_{xx}(t)$ is found to 
fit very well a single exponential of the form $R_{xx}(t)=\alpha+\beta e^{-t/T_{1\prime}}$ (solid line in the inset of Fig.2).
Here, $T_{1\prime}$ is the characteristic relaxation time of the resistance and $\alpha, \beta$
are coefficients which determine the on-resonance and off-resonance resistance values. At temperatures
$T\gtrsim 30$ mK, the {\it maximum} change in the Zeeman gap 
between on-resonance and off-resonance is $(\delta\Delta_{z})^{max}=g^{\star}\mu_{B} B_{N}$ 
which is smaller  than $2k_{B}T$ by at least a factor of 4.  In addition,  the $T_{1}$
measurements  were performed in the small-power limit such that
$(\delta R_{xx})_{low-power}\lesssim 0.1\cdot(\delta R_{xx})_{max}$,
i.e. with partially depolarized nuclei only. Therefore, $\delta\Delta_{z}\sim \delta B_{N}\ll 2k_{B}T$, and to first order
the resistance scales as   $\delta R_{xx}\propto \frac{g^{\star}\mu_{B} \delta B_{N}}{2k_{B}T}$. Noting that 
$B_{N}\propto {\cal M}$,  we find $T_{1\prime}\simeq T_{1}$ to a very good approximation.

The main panel of Fig.2 shows the temperature dependence of the spin-lattice 
relaxation times $T_{1}$ for the $^{75}$As nuclei at filling factors $\nu=0.84$ (diamonds),
0.86 (filled circles) and 0.895 (empty circles). The temperature quoted
is to a very good approximation the actual electronic temperature, $T_{e}$.
This was determined by using
the electronic resistance as an {\it in situ} thermometer.
The x-axis error bars are remaining uncertainties in determining this temperature. Each
$T_{1}$ datum was reproduced over at least three independent measurements.
The y-axis error bar provides a range for the scattering at each data point. 
No systematic dependence of $T_{1}$ on the RF-power  was observed.

 \begin{figure}
\begin{center}
\includegraphics[width=7.5cm]{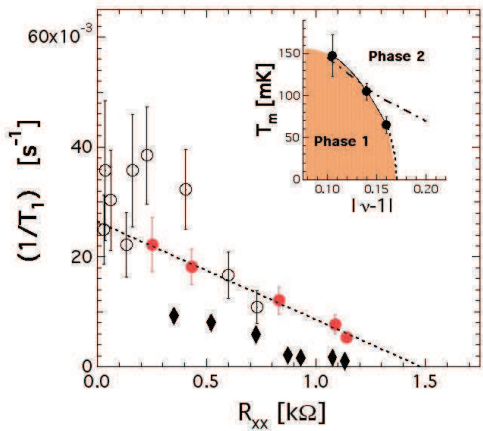}
\end{center}
\vspace*{-5mm}\caption{  {\small Nuclear spin-lattice relaxation rate $(1/T_{1})$ for $^{75}As$ versus $R_{xx}$ at the same filling factors as
in Fig. 2.  The straight line is a linear fit to the $\nu=0.86$ data.
$T_{m}$ (defined in the text) is shown 
in the inset (filled circles) versus the partial filling factor, $|\nu-1|$. The solid line is a guide-to-the-eye, 
and the dotted line an extrapolation to T=0. 
The dash-dotted line is a theoretical estimate for the classical
transition temperature of a Skyrme crystal \cite{Cote97}(see text).}}
\label{fig3}
\end{figure}

The relaxation time  decreases significantly as $T\rightarrow 0$ for the three filling factors. The 
data at $\nu=0.895$ 
shows a similar behaviour near $T\sim 130$ mK, at which point we could no longer 
detect  the NMR signal. These data show that  the nuclear
spins relax much more efficiently as $T\rightarrow 0$
and that strong magnetic fluctuations exist in the quantum many-body 
 ground state. Interestingly, $T_{1}$ also becomes shorter with  $\nu$ increasing. 
 Barrett {\it et al.} showed
 that the nuclear spin relaxation was strongly suppressed for filling factor at, or very close to 
 $\nu=1$ and at $T\sim 1$K \cite{Barrett95,Tycko95} . Our data are outside of this regime, $|\nu-1|\lesssim 0.1$, 
 and it is difficult to address this filling factor region with resistive NMR since the resistance vanishes in the quantum Hall state. Nevertheless, we expect $T_{1}$ to become
 much longer when the first Landau level is completely filled and
 the resulting ferromagnetic state suppresses the spin degree-of-freedom.
 
Inspecting the data of Fig.2 one wonders as to the origin of this strong T-dependence of
$T_{1}$. In fact, given that $R_{xx}$ itself is strongly T-dependent and $\nu$-dependent 
the relationship seen in Fig.2 may reflect a correlation with $R_{xx}$. Towards
this end we note that in the quantum Hall
regime vanishing $R_{xx}$ implies 
vanishing conductivity owing to the tensor inversion, $\sigma_{xx}=\rho_{xx}/(\rho_{xx}^{2}+\rho_{xy}^{2})$,
and a finite quantized diagonal resistivity, $\rho_{xy}$. This insulating behavior 
in the quantum Hall regime is a consequence of the localization of the electrons
(holes)  by disorder with density $\nu^{\ast}=\nu-K$ ($<0$ for holes). Here, $K$
is an integer which equals one at $\nu=1$.  
Recent micro-wave measurements
performed on similar high-mobility samples show pinning resonances 
which suggest that a collective electron solid is formed in this regime \cite{Chen03}.

 Figure 3 shows
 the spin-lattice relaxation rate $(1/T_{1})$ of Fig.2 plotted versus $R_{xx}$. 
It shows a remarkable linear relationship linking an increase in relaxation rate
with decreasing $R_{xx}$ and hence with increasing localization. 
 This strongly suggests
 that the nuclear spin relaxation is induced predominantly by those electrons 
 (or holes) forming
 a solid, rather than the remaining conducting electronic states contributing to $\sigma_{xx}$. 
 It is important to realize that
 the data follow a trend {\it opposite} to the usual Korringa behavior observed in metals, 
  $(1/T_{1}) \propto N_{F} \propto \sigma_{xx}\propto \rho_{xx}$ ($N_{F}$ is the electronic density
  of states). 
  If the Korringa relation were to hold, one would expect stronger relaxation at higher
  values of $R_{xx}$, opposite to our data. 
  
Efficient relaxation of the nuclear spins requires magnetic fluctuations in their environment. 
The data in Fig.3 requires an increase of such fluctuations as $R_{xx}$ (and hence $\sigma_{xx}$)
decreases. A spin-polarized two-dimensional Fermi gas is extremely inefficient in 
providing such fluctuations. For instance, 
we measured the spin-lattice relaxation time $T_{1}$ in the high-field phase ($B\gtrsim 30T$) at small filling factors
$\nu<\frac{2}{9}$ and $\nu<\frac{1}{5}$ where a Wigner crystal phase of electrons is expected to 
form\cite{Gervais04}.
In contrast to near $\nu\sim 1$,  $T_{1}$ in the Wigner crystal regime was found to be 
very long, ranging from $\sim$350 to  1000 s at $T\sim 50$ mK and for all values of $R_{xx}$.
On the other hand, a spin-wave Goldstone mode of a Skyrme crystal \cite{Cote97}
provides a very efficient mechanism for relaxing the nuclear spins.
At magnetic fields $H_{0}\sim  7T$ near $\nu\sim 1$, our sample has a Zeeman-to-Coulomb energy
ratio $\eta \simeq 0.015$ which favors Skyrmion formation and well below the 
 $\eta_{c}\simeq 0.022$ where they are expected to disappear \cite{Fertig94}. 
Such a Skyrmion crystal was calculated to enhance the nuclear spin relaxation
by a factor $\sim10^{3}$ over that of a 2D Fermi gas. This is consistent with the  $\sim10^{1}-10^{2}$ increase 
in relaxation that we observed near $\nu\sim 1$ as compared to the rate in
the high-field electron solid phase. In addition, we have measured $T_{1}$ in the same sample
around $\nu=3$, and found that $T_{1}\gtrsim 300$ s for $T\rightarrow 0$ and $R_{xx}\rightarrow 0$
 in contrast to our result near $\nu\sim 1$. This is consistent with the result 
 by Schmeller {\it et al.}\cite{Schmeller95} which showed no Skyrmion formation
 to occur at  $\nu\sim 3$.

Extrapolating the rate in Fig.3 to the x-axis defines a resistance $R_{m}$ at 
which $T_{1}\rightarrow \infty$. While there will remain other, weaker relaxation mechanisms, this
extrapolated $R_{m}$ should provide a measure where nuclear relaxation by the strong 
low-temperature mechanism ceases. When translated to $T_{m}$ via our $R_{xx}$ versus T
calibration, the $T_{m}$'s  define a temperature boundary for the low-temperature 
magnetic phase as shown in the inset 
of Fig.3 versus the partial filling factor, $|\nu-1|$.  The error bars correspond to 
errors in determining $R_{m}$ from
extrapolating to $(1/T_{1})\rightarrow 0$. The solid line in the inset 
is a guide-to-the-eye and the dotted line is 
an extrapolation to zero temperature. The shaded region is a proposed partial phase diagram spanned 
by the low-temperature magnetic phase (phase 1). The curvature of the data
suggests the existence of a critical filling factor $\nu_{m}\sim 0.83$ defining a
quantum phase transition between phases 1 and 2. A similar phase diagram
has been deduced in the heat capacity work of Bayot {\it et al.}\cite{Bayot96,Bayot97}.
While in their work the overall values of $T_{m}$ are smaller, we
expect the transition to be sensitive to disorder so the discrepancy 
may simply reflect the overall lower disorder in our sample.

An estimate for the classical melting temperature of a Skyrme crystal can be 
made from the relation $T_{crystal}=(b^{2}\mu/4\pi)$. The shear modulus $\mu$
of the Skyrme crystal can be calculated from a microscopic theory, 
which together with the lattice constant $b$
yield $T_{crystal}\sim 0.363K $  at $|\nu-1|\sim 0.85$ \cite{Cote97}. Similar calculations for 
the melting of an electron crystal usually overestimate the melting transition by a factor of 2, and quantum 
positional fluctuations are expected to further suppress it. Using the value 
for $T_{crystal}$ given in Ref.\cite{Cote97}, we
plotted its partial filling factor dependence  in the inset of Fig.3 
with a dash-dotted line. Its magnitude was rescaled by a factor of 3.6 
so as to match the overall magnitude of our data. 
While the trend of the  data is not completely accounted for, the model, indeed, predicts
$T_{m}$ decreasing with increasing $|\nu-1|$. Within this scenario, phase 1 would correspond to 
a square lattice phase of Skyrmions with long-range positional and orientation order, while
phase 2 would correspond possibly to a melted skyrmion phase with quasi long-range magnetic order only.

In conclusion, we presented a detailed study of the nuclear spin-lattice relaxation rate 
near $\nu=1$ in the $T\rightarrow 0$ limit of an extremely high-quality GaAs/AlGaAs sample.
In the vicinity of the $\nu=1$ quantum Hall state, the nuclear spin relaxation increases
strongly as the temperature is lowered and when $R_{xx}\rightarrow 0$. This 
strongly suggests that the localized states are responsible for the  fast nuclear relaxation.
 We find a natural interpretation of our data in terms of a magnetic phase of localized 
 skyrmions relaxing the nuclear spin via a Goldstone mode of the crystal and deduce
 a partial phase diagram in the $T-\nu$ plane. 

The authors would like to acknowledge helpful discussions with Yong Chen, Lloyd Engel, 
Herb Fertig, Michael Hilke, Ren\'{e} C\^{o}t\'e, Mansour Shayegan
and Kun Yang.  One of the authors (G.G.) is grateful for the hospitality of the National
High-Magnetic Field Laboratory. Research funded by the NSF 
under grant \# DMR-0084173 and \# DMR-03-52738, and by the DOE
under grant \# DE-AIO2-04ER46133.

\end{document}